\documentclass[pre,aps ,superscriptaddress,showpacs]{revtex4-1}
\usepackage{natbib}
\usepackage{amsmath}
\usepackage{epsfig}
 \usepackage{color}

\begin{document}

\title{A fluctuation theorem for Floquet quantum master equations}
\author{Fei Liu}
\email[Email address: ]{feiliu@buaa.edu.cn}
\affiliation{School of Physics, Beihang University, Beijing 100191, China}

\date{\today}

\begin{abstract}
{We present a fluctuation theorem for Floquet quantum master equations. This is a detailed version of the famous Gallavotti-Cohen theorem. In contrast to the latter theorem, which involves the probability distribution of the total heat current, the former involves the joint probability distribution of positive and negative heat currents and can be used to derive the latter. A quantum two-level system driven by a periodic external field is used to verify this result.}

\end{abstract}
%\pacs{05.70. Ln, 05.30.-d}
\maketitle

\section{Introduction}
\label{section1}
The Gallavotti-Cohen (GC) fluctuation theorem states that the ratio of the probability distribution $p(\sigma)$ of having an average total entropy production rate $\sigma$ to that of having $-\sigma$ approaches $\exp(t\sigma)$ as the time interval $t$ increases; that is,
% Editor: Please ensure that the intended meaning has been maintained in the edits of the previous sentence.
\begin{eqnarray}
\label{orgGC}
\frac{p(\sigma)}{p(-\sigma ) }\asymp e^{t \sigma},
\end{eqnarray}
where the Boltzmann constant $k_B$ is set to 1 throughout this paper, and the symbol $\asymp$ denotes asymptotic change as $t$$\rightarrow$$ \infty$~\cite{Touchette2008}. The original GC fluctuation theorem was inspired by a relationship between the probabilities of fluctuations in the shear stress of fluids in nonequilibrium steady states~\cite{Evans1993} and was proved in modern dynamical system theory~\cite{Gallavotti1995,Gallavotti1995a}. Complicated techniques were used; see the latest review~\cite{Gallavotti2019}. In contrast, its proof in stochastic systems, e.g., in classical Langevin systems or discrete jump systems, is simple~\cite{Kurchan1998,Lebowitz1999,Maes1999,Esposito2009}. Hence, stochastic dynamical systems are suitable to explore new fluctuation theorems.

The motivation of this paper is as follows. Let us imagine that a system contacts a heat bath having an inverse temperature $\beta$ and is in a nonequilibrium steady state due to some external force. The GC theorem can be reexpressed in terms of the probability distribution of the total heat current $j$, the total released heat averaged over the time interval $t$. Then, the exponent on the right-hand side of Eq.~(\ref{orgGC}) is replaced by $t\beta j$. We know that the total heat is composed of positive and negative parts. Accordingly, the total heat current $j$ can be divided into $j_+$ and $j_-$. Apparently, the individual current
does not satisfy the GC fluctuation theorem. However, is the same true of the joined currents? Here, a positive answer is presented, at least for the driven quantum systems described by Floquet quantum master equations~\cite{Grifoni1998,Kohn2001,Szczygielski2013,Gasparinetti2014,Cuetara2015}. %In addition to having a microscopic physical foundation, the major advantage of applying these systems is that they have nonequilibrium steady-state solutions even if the systems contact only one heat bath~\cite{Szczygielski2013}. This is essential to make our result relevant to the GC fluctuation theorem.

The rest of this paper is organized as follows. In Sec.~(\ref{section2}), we review the Floquet quantum master equation and its stochastic thermodynamics. In Sec.~(\ref{section3}), we prove a fluctuation theorem. In Sec.~(\ref{section4}), a two-level quantum system is used to concretely verify this theorem. Section~(\ref{section5}) concludes the paper.

\section{A Floquet quantum master equation}
\label{section2}
Given the Hamiltonian of a quantum system driven by periodic external forces, denoted as $H(t)$, we have
\begin{eqnarray}
H\left(t+{\cal T}\right)=H(t),
\end{eqnarray}
where ${\cal T}$$=$$2\pi/\Omega$ is the periodicity and $\Omega$ is the driving frequency. According to the Floquet theorem~\cite{Zeldovich1967,Shirley1965}, this periodic Hamiltonian satisfies an eigenvalue equation:
\begin{eqnarray}
(H(t)-i\partial_t )|u_n(t)\rangle =\epsilon_n|u_n(t)\rangle,
\end{eqnarray}
where $\epsilon_n$ and $|u_n(t)\rangle$ ($n$$=$$1,\cdots,N$) are quasi-energies and Floquet bases, respectively, and we set $\hbar$$=$$1$. Note that the Floquet bases are orthonormal and periodic. In addition, we emphasize that these quasi-energies are restricted in a zone with a size of $\Omega$. The heat bath that interacts with the quantum system has an inverse temperature $\beta$. Under the weak system-bath coupling conditions and time scale separation assumptions, the evolution of the reduced density matrix of the quantum system $\rho(t)$ can be described by the Floquet quantum master equation~\cite{Grifoni1998,Breuer1997,Alicki2006}:
\begin{eqnarray}
\label{FQME}
\partial_{t}\rho(t)=-i[H(t),\rho(t)]+D(t)[\rho(t)].
\end{eqnarray}
The $D$-term represents the dissipation induced by the interaction between the system and the heat bath and is
\begin{eqnarray}
\label{dissipations}
&&D(t)[\rho]\nonumber \\
&=&\sum_{\omega} r(\omega )\left[A(\omega ,t)\rho A^\dag(\omega ,t)- \frac{1}{2}
\left\{A^\dag (\omega ,t)A(\omega ,t),\rho\right\}\right].
\end{eqnarray}
In the above equation, $\omega$ are Bohr frequencies and are equal to $\epsilon_n-\epsilon_m + q\Omega$, where $q$ are certain integer numbers. The numbers may be positive or negative but always appear in pairs. Additionally, in the same equation, $A(\omega,t)$ and $A^\dag(\omega,t)$ are called the Lindblad operators and are related by
\begin{eqnarray}
\label{Hermitiancondition}
A(-\omega,t)=A^\dag(\omega,t).
\end{eqnarray}
The interaction operator of the quantum system and the heat bath is given as $A$$\otimes$$B$, where $A$ and $B$ are the system and heat bath components, respectively. These Lindblad operators are obtained by performing a Fourier-like expansion of the interaction-picture operator of $A$~\cite{Breuer1997}:
\begin{eqnarray}
\label{Amatrix}
%U^\dag(t,t_0)AU(t,t_0)=\sum_{m,n,q}  A(\omega_{mn}^{q},t_0)\exp[ -i(t-t_0)\omega_{mn}^q ]\nonumber \\
A(\omega,t)&=&\sum_{m,n,q}\delta_{\omega,\epsilon_n-\epsilon_m + q\Omega}\langle\langle u_m|A|u_n\rangle \rangle_q |u_m(t)\rangle\langle u_n(t)|e^{-iq\Omega t},
\end{eqnarray}
where $\delta$ is the Kronecker symbol, and the time-independent coefficient $\langle\langle u_m|A|u_n\rangle \rangle_q $ is the $q$-th harmonic of the transition amplitude $\langle u_m(t)|A|u_n(t) \rangle$; that is,
\begin{eqnarray}
\langle\langle u_m|A|u_n\rangle \rangle_q =\frac{1}{\cal T}\int_0^{\cal T} \langle u_m(t)|A|u_n(t) \rangle e^{iq\Omega t}.
\end{eqnarray}
The last ingredient of the Floquet quantum master equation is the assumption that the heat bath is always in a thermal state. Then, $r(\omega)$, the Fourier transformation of the correlation function of the operator $B$, satisfies the important Kubo-Martin-Schwinger (KMS) condition~\cite{Breuer2000,Rivas2012}:
\begin{eqnarray}
\label{KMScondition}
r(-\omega )=r(\omega )e^{ -\beta \omega}.
\end{eqnarray}

\section{A fluctuation theorem}
\label{section3}
Stochastic thermodynamics can be established for the Floquet quantum master equation~\cite{Liu2016a,Liu2018,Cuetara2015}. Roughly, Eq.~(\ref{FQME}) is unraveled into the dynamics of individual quantum systems~\cite{Carmichael1993,Plenio1998,Breuer2002,Wiseman2010}. The evolution of each system is composed of a continuous process alternating with discrete random jumps.
% Quality control editor: Please ensure that the intended meaning has been maintained in the edits of the previous sentence.
Assume that these jumps occur at time points $t_i$ ($i$$=$$1,\cdots$). Each jump indicates that a quantum $\omega_i$ is released to the heat bath~\cite{Breuer2003,DeRoeck2006,Crooks2008,Horowitz2012,Hekking2013,Liu2016a,Liu2018}. The subscript $i$ represents the time points of these energy exchanges. When the evolution of a quantum system ends at time $t$, a quantum jump trajectory is generated and is marked as $\{\overrightarrow \omega\}$$=$$\{\omega_1,\cdots\}$. If the density matrixes of these individual quantum systems are $\widetilde\rho(\{\overrightarrow \omega\},t)$, their average weighted by the probabilities of all possible quantum jump trajectories is just the reduced density matrix $\rho(t)$ of Eq.~(\ref{FQME}). From a thermodynamic point of view, the quanta are the heat released to the heat bath. Hence, given a quantum jump trajectory $\overrightarrow\omega$, we define the total heat along it as
\begin{eqnarray}
Q\{\overrightarrow \omega\}=\sum_{i=1}\omega_{i}=Q_+\{\overrightarrow \omega \}+Q_-\{\overrightarrow \omega \}.
\end{eqnarray}
In the second equation, we specifically define the positive and negative heat, $Q_+\{\vec\omega \}$ and $Q_-\{\vec\omega \}$, respectively; clearly, they are simply equal to the sums of positive and negative Bohr frequencies.

Because the occurrences of quantum jump trajectories and time points of quantum jumps are random events, all three types of heat are stochastic quantities. Let the joint probability distribution of the positive and negative heat, $Q_+$ and $Q_-$, respectively, be $p(Q_+,Q_-)$. We can construct its histogram by directly simulating quantum jump trajectories~\cite{Liu2016a,Liu2018}. Because we are interested in the statistics over long time limits, a more practical approach is to compute its moment generation function,
\begin{eqnarray}
\label{heatCF}
\Phi (\chi_+,\chi_-)&=&\int dQ_+ dQ_- p(Q_+,Q_-) e^{\chi_+Q_+ +\chi_-Q_- } \nonumber \\
&=&{\rm Tr}[\hat\rho(t)].
\end{eqnarray}
We introduce a characteristic operator $\hat\rho(t)$ in the second equation above and find that it satisfies an evolution equation:
\begin{eqnarray}
\label{equationofmotionforrhohat}
\partial_t \hat{\rho}(t)&=&-i[H(t), \hat{{\rho}}]+D(t,\chi_+,\chi_-)[\hat{\rho}(t)],
\end{eqnarray}
where the super-operator
\begin{eqnarray}
\label{modifieddissipation}
&&D(t,\chi_+,\chi_-)[\hat\rho ]\nonumber \\
&=&\sum_{\omega>0 }  r(\omega)\left[ e^{i\chi_+\omega} A(\omega,t)\hat\rho
A^\dag(\omega,t)- \frac{1}{2}
\left\{A^\dag (\omega,t)A(\omega,t),\hat\rho \right\}\right]+\nonumber \\
&&\sum_{\omega<0 } r(\omega)\left[ e^{i\chi_-\omega} A(\omega_,t)\hat\rho  A^\dag(\omega,t)- \frac{1}{2}
\left\{A^\dag (\omega,t)A(\omega,t),\hat\rho \right\}\right]. %+ \nonumber \\
%&&\sum_{\omega>0}r_k(-\omega)\left[A_k^\dag(\omega,t)O A_k(\omega,t)- \frac{1}{2}
%\left\{A_k(\omega,t)A_k^\dag(\omega,t),O\right\}\right].
\end{eqnarray}
If $\chi_\pm$$=$$0$, Eq.~(\ref{modifieddissipation}) reduces to Eq.~(\ref{FQME}). This result comes from a simple extension of the previous equation (Eq. (19) in Ref.~\cite{Liu2016a}), which concerned the moment-generating function of the total heat, and we can reobtain it by letting $\chi_+$$=$$\chi_-$ in Eq.~(\ref{modifieddissipation}). Because the derivation is the same, we do not repeat it here.

The abstract Eq.~(\ref{equationofmotionforrhohat}) is not the most convenient to use in analyses. According to Eq.~(\ref{heatCF}), $\Phi (\chi_+,\chi_-)$ is equal to a sum of diagonal elements of $\hat\rho(t)$. Hence, we write the evolution equations for $P_n(t)$$=$$\langle u_n(t)|\hat\rho(t)|u_n(t)\rangle $ in the Floquet bases:
\begin{eqnarray}
\frac{d}{dt} {\textbf P}(t)=\textbf{R}(\lambda_+,\lambda_-) {\textbf P}(t),
\end{eqnarray}
where the vectors ${\textbf P}(t)$$=$$(P_1(t),\cdots,P_N(t))^T$ and $T$ represents the transpose and the nondiagonal matrix elements
\begin{eqnarray}
&&[{\textbf R}(\lambda_+,\lambda_-)]_{nm}\nonumber \\
&=&\sum_{\omega>0} e^{\lambda_+\omega} r(\omega)|\langle u_n(t)|A(\omega,t)|u_m(t)\rangle|^2 +\nonumber  \\
   &&\sum_{\omega<0} e^{\lambda_-\omega} r(\omega)|\langle u_n(t)|A(\omega,t)|u_m(t)\rangle|^2,
%\left\{
 % \begin{array}{ll}
  %  \sum_{\omega>0} e^{\lambda_+\omega} r(\omega)|\langle u_n(t)|A(t)|u_m(t)\rangle|^2 , & \hbox{1;} \\
  %  \sum_{\omega<0} e^{\lambda_-\omega} r(\omega)|\langle u_n(t)|A(t)|u_m(t)\rangle|^2 , & \hbox{1.}
  %\end{array}
%\right.
\end{eqnarray}
($m\neq n$), and the diagonal elements
\begin{eqnarray}
&&[{\textbf R}(\lambda_+,\lambda_-)]_{nn}\nonumber \\
&=& \sum_{\omega>0} e^{\lambda_+\omega}  r(\omega=q\Omega)|\langle u_n(t)|A(\omega,t)|u_n(t)\rangle|^2
+ \nonumber \\
&& \sum_{\omega<0}e^{\lambda_-\omega} r(\omega=q\Omega)|\langle u_n(t)|A(\omega,t)|u_n(t)\rangle|^2 \nonumber - \\
&& \sum_\omega r(\omega) \sum_m |\langle u_m(t)|A(\omega,t)|u_n(t)\rangle|^2.
\end{eqnarray}
Note that Eq.~(\ref{Amatrix}) reminds us that $\textbf{R}(\lambda_+,\lambda_-) $ is in fact a constant matrix. We can easily prove that this matrix possesses symmetry:
\begin{eqnarray}
\label{matrixAsymmetry}
\left[\textbf{R}(\chi_+,\chi_-)\right]^T=\textbf{R}(-\beta-\chi_-,-\beta-\chi_+).
\end{eqnarray}
Using this property, we obtain a fluctuation theorem by simply following a standard procedure, e.g., that presented by Lebowitz and Spohn~\cite{Lebowitz1999}. First, because the transpose matrix has the same eigenvalues as the original matrix, Eq.~(\ref{matrixAsymmetry}) implies that the scaled cumulant generating function~\cite{Touchette2008}, $\phi(\chi_+,\chi_-)$, or the maximal eigenvalue of the $\textbf R$-matrix
%\begin{eqnarray}
%\phi(\chi_+,\chi_-)=\lim_{t\rightarrow \infty}\frac{1}{t}\ln \Phi(\chi_-,\chi_+),
%\end{eqnarray}
has the same symmetry:
\begin{eqnarray}
\label{symmetricSCGF}
\phi(\chi_+,\chi_-)= \phi(-\beta-\chi_-,-\beta-\chi_+).
\end{eqnarray}
Given the large deviation function $I(j_+,j_-)$ of the distribution $p(j_+,j_-)$ for the positive heat current $j_+$$=$$Q_+/t$ and the negative heat current $j_-$$=$$Q_-/t$, and because the function is a Legendre transform of the scaled cumulant generating function,
%\begin{eqnarray}
%I(j_+,j_-)=\max_{\lambda_+,\lambda_-}[ \lambda_+j_+ +\lambda_-j_- -\phi(\lambda_+,\lambda_-)]
%\end{eqnarray}
Eq.~(\ref{symmetricSCGF}) immediately leads to
\begin{eqnarray}
\label{largedeviationfunctionsymmetry}
I(j_+,j_-)=I(-j_-,-j_+) -\beta (j_++j_-).
\end{eqnarray}
Then, the probability distribution for these two heat currents satisfies the fluctuation theorem
\begin{eqnarray}
\label{extendedGCT}
\frac{p(j_+,j_-)}{p(-j_-,-j_+) }\asymp e^{t\beta (j_++j_-) }.
\end{eqnarray}
The conventional GC fluctuation theorem~(\ref{orgGC}) can be easily derived from Eq.~(\ref{extendedGCT}).

Fluctuation theorem~(\ref{extendedGCT}) has a time-reversal explanation. Analogous to that of classical stochastic processes~\cite{Lebowitz1999}, the ratio of the probability distribution ${\cal P}\{\overrightarrow\omega\}$ of observing a quantum jump trajectory $\overrightarrow \omega$$=$$\{\omega_1,\cdots,\omega_M\}$ to ${\cal P}\{\overleftarrow\omega\}$ of observing its reversed trajectory $\overleftarrow\omega$$=$$\{-\omega_M,\cdots,-\omega_1\}$ approaches $\exp(\beta Q\{\overrightarrow\omega\})$ as the time interval $t$ increases~\cite{Liu2018}. Note that not only is the time order of the quantum jumps reversed in the revered trajectory, but the signs of these Bohr frequencies are reversed, we obtain $Q_\pm\{\overleftarrow \omega\}$$=$$-Q_\mp\{\overrightarrow \omega\}$ and
\begin{eqnarray}
p(j_+,j_-)&= &\int {\cal D}\overrightarrow \omega {\cal P}\{\overrightarrow\omega\} \delta(Q_+\{\overrightarrow\omega\}-Q_+)\delta(Q_-\{\overrightarrow\omega-Q_-\})\nonumber\\
&\asymp&\int {\cal D}\overleftarrow \omega {\cal P} \{\overleftarrow\omega\}e^{-\beta(Q_-\{\overleftarrow\omega\}+Q_+\{\overleftarrow\omega\}) } \delta(Q_+\{\overleftarrow\omega\}+Q_-)\delta(Q_-\{\overleftarrow\omega\}+Q_+)\nonumber\\
&= &e^{t\beta(j_+ +j_- ) } p(-j_-,-j_+).
\end{eqnarray}
Some unimportant constants are ignored here. This rough proof explains why the positions of $j_+$ and $j_-$ in the probability distribution on the right-hand side are exchanged and minus signs are added simultaneously; they are just the consequences of the time-reversal. Eqs.~(\ref{largedeviationfunctionsymmetry}) and~(\ref{extendedGCT}) are the central results of this paper. In the next section, we use a two-level quantum system to concretely show these results.

\section{Two-level quantum system}
\label{section4}
Consider the Hamiltonian of a two-level quantum system~\cite{Breuer1997,Szczygielski2013,Langemeyer2014,Gasparinetti2014,Cuetara2015}
\begin{eqnarray}
\label{TLS}
H(t)=\frac{1}{2} {\omega_0}\sigma_z +\frac{1}{2}{\Omega_R}\left(\sigma_+ e^{-i\Omega t}+\sigma_- e^{i\Omega t}\right),
\end{eqnarray}
where $\omega_0$ is the transition frequency of the bare system, $\Omega_R$ is the Rabi frequency, and $\Omega$ is the frequency of the periodic external field. The Floquet bases and the quasi-energies of this system are
\begin{eqnarray}
\label{TLSFloquetbases}
|u_{\pm}(t)\rangle =\frac{1}{\sqrt{2\Omega'}}
\left(\begin{array}{c}
 \pm \sqrt{\Omega'\pm\delta}\\
 e^{i\Omega t}\sqrt{\Omega'\mp\delta},
\end{array}\right),
\end{eqnarray}
and $\epsilon_\pm$$=$$(\Omega \pm \Omega')/2$, respectively, where $\Omega'$$=$$\sqrt{\delta ^2+\Omega_R^2}$ and the detuning parameter $\delta$$=$$\omega_0$$-$$\Omega$. Here, we additionally set $\Omega$$>$$\Omega'$. %~\footnote{If this condition is not satisfied, the matrixes $(\textbf{R})_{12}$ and $(\textbf{R})_{21}$ in Eq.~(\ref{twolevelmatrixes}) need to be slightly changed. But it does not affect our conclusion.}.
Assume that the coupling between the two-level system and the heat bath is $\sigma_x$-coupling. There are six Lindblad operators, and the Bohr frequencies are $\pm\Omega $, $\pm(\Omega-\Omega')$, and $\pm(\Omega+\Omega')$. By performing some simple derivations, the ${\textbf R}$-matrix is obtained:
\begin{eqnarray}
\label{twolevelmatrixes}
(\textbf{R})_{11}&=& (e^{i\chi_+\Omega}-1)\Gamma_{+\Omega}+(e^{-i\chi_-\Omega}-1)\Gamma_{-\Omega}-\Gamma_{-(\Omega-\Omega')}-\Gamma_{+(\Omega+\Omega')}, \nonumber\\
(\textbf{R})_{12}&=&e^{ i\chi_+(\Omega-\Omega')}\Gamma_{+(\Omega-\Omega')} + e^{- i\chi_-(\Omega+\Omega')}\Gamma_{ -(\Omega+\Omega')}, \nonumber \\
(\textbf{R})_{21}&=&e^{-i\chi_-(\Omega-\Omega')}\Gamma_{-(\Omega-\Omega')}+ e^{i\chi_+(\Omega+\Omega')}\Gamma_{ +(\Omega+\Omega')},\nonumber \\
(\textbf{R})_{22}&=&(e^{i\chi_+\Omega}-1)\Gamma_{+\Omega}  + (e^{-i\chi_1^-\Omega}-1)\Gamma_{ -\Omega} -\Gamma_{+(\Omega -\Omega')}-
\Gamma_{-(\Omega+\Omega')},
\end{eqnarray}
where we do not explicitly write $\chi_\pm$ on the left-hand side and the coefficients are
\begin{eqnarray}
\label{coeffs}
\Gamma_{\pm\Omega}&=&\left(\frac{\Omega_R}{2\Omega'}\right)^2 r(\pm\Omega), \nonumber \\
\Gamma_{\pm(\Omega-\Omega')} &=&\left(\frac{\delta-\Omega_R}{2\Omega'}\right)^2 r(\pm(\Omega-\Omega')), \\
\Gamma_{\pm(\Omega+\Omega')}&=&\left(\frac{\delta+\Omega_R}{2\Omega'}\right)^2 r(\pm(\Omega+\Omega')).\nonumber
\end{eqnarray}
In the matrix, symmetry~(\ref{matrixAsymmetry}) is apparent. Because this is a simple $2$$\times$$2$ matrix, the analytical expression of the scaled cumulant generating function is
\begin{eqnarray}
\label{scaledCGFtwolevel}
\phi(\chi_+,\chi_-)&=&\frac{1}{2}[({\textbf R})_{11}+ ({\textbf R})_{22}\pm B] ,
\end{eqnarray}
where
\begin{eqnarray}
B&=&\sqrt{[({\textbf R})_{11}-({\textbf R})_{22}]^2+4({\textbf R})_{12}({\textbf R})_{21}}.
\end{eqnarray}
We clearly see that symmetry~(\ref{matrixAsymmetry}) is inherited in Eq.~(\ref{scaledCGFtwolevel}).

\begin{figure}
\includegraphics[width=1.\columnwidth]{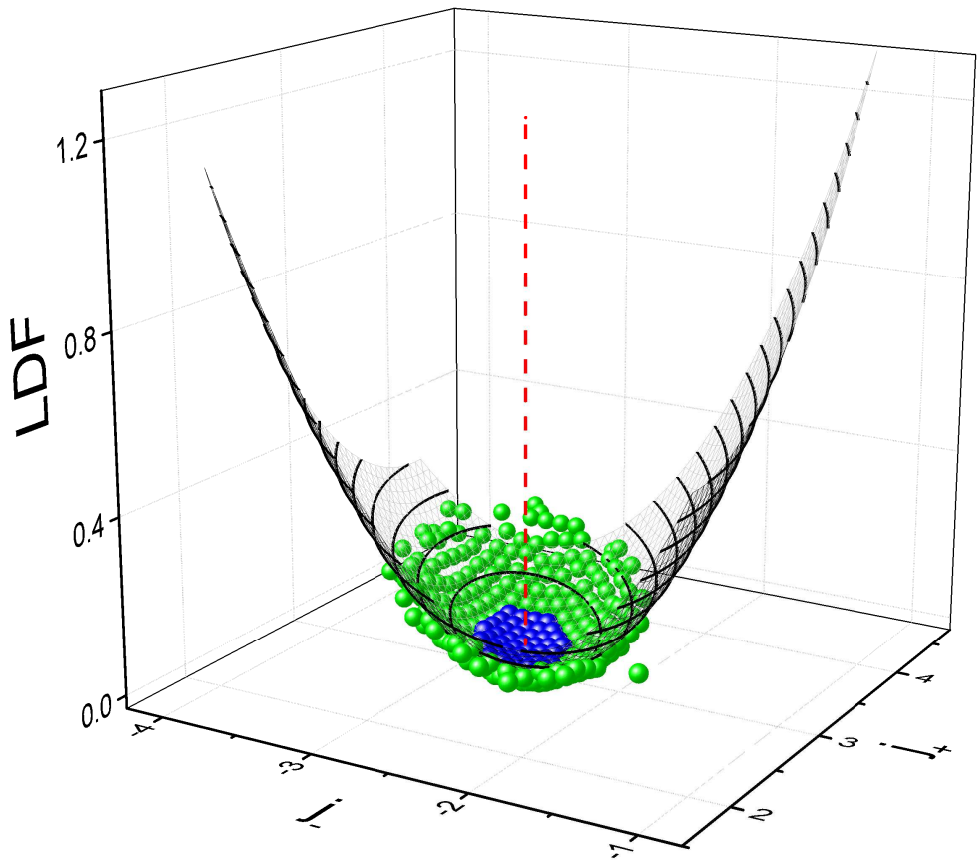}
\caption{The large deviation functions (LDF) $I(j_+,j_-)$ for a two-level Floquet quantum system. The meshed surface is obtained by performing a Legendre transformation on Eq.~(\ref{scaledCGFtwolevel}). The spheres are the data of simulating quantum jump trajectories, where the simulation times for the green and blue data are 50 and 500, respectively. The Fourier transformation of the correlation function is set to be $r(\omega)$$=$${\cal A}|{\omega}|^3 {\cal N}_k(\omega)$ for $\omega$$<$$0$; otherwise, $r(\omega)$$=$${\cal A}|{\omega}|^3 [{\cal N}_k(\omega)+1]$, where ${\cal N}(\omega)$$=$$1/[\exp ( \beta |\omega| )-1]$, and the coefficient $A$ is related to the coupling strength between the system and the heat bath~\cite{Breuer2000}. The parameters used are $\omega_0$$=$$1$, $\Omega_R$$=$$0.8$, $\Omega$$=$$1.1$, ${\cal A}$$=$$1$, and $\beta$$=$$1/3$. The red dashed line indicates the location of the two mean currents.}
\label{fig1}
\end{figure}

We compute the large deviation function $I(j_+,j_-)$ by numerically performing a Legendre transformation on Eq.~(\ref{scaledCGFtwolevel}) and depict it in Fig.~(\ref{fig1}). Eq.~(\ref{largedeviationfunctionsymmetry}) can be easily verified (data not shown). On the other hand, it is interesting to approximately compute the large deviation function by simulating quantum jump trajectories with finite times; see
% Editor: On the line below, please ensure that the intended meaning has been maintained in this edit here and elsewhere throughout the manuscript.
the spheres
in the same figure. We find that even if the simulation time is short, the simulated data roughly exhibit the profile of the function. Although longer simulation times lead to better results, the regimes of the sampled data decrease around the mean currents $(\overline{j}_+,\overline {j}_-)$. This observation reminds us that simply simulating quantum jump trajectories is not enough to solve a large deviation function.

\iffalse
Finally, we attempt to verify the fluctuation theorem~(\ref{extendedGCT}) by computing the function
\begin{eqnarray}
\label{deltafunction}
\Delta(t)=\frac{1}{t}\ln\frac{p(j_+,j_-)}{p(-j_-,-j_+)} -\beta(j_++j_-)
\end{eqnarray}
and checking its degree of deviation from zero at finite simulation times; see Fig~(\ref{fig2}).
\begin{figure}
%\includegraphics[width=1.\columnwidth]{figure2.eps}
\caption{The simulation data for Eq.~(\ref{deltafunction}). The red and blue bars are the data of simulation times 10 and 100, respectively. Note that the dark red and blue bars denote that these data are negative. The other parameters are the same as those used in Fig.~(\ref{fig1}). }
\label{fig2}
\end{figure}
We see that although the data with longer simulation times almost overlap with the $j_+$-$j_-$ plane, indicating that the fluctuation theorem tends to be true, they sample a very limited regime around the mean currents.
% Editor: Please ensure that the intended meaning has been maintained in the edits of the previous sentence.
\fi

\section{Conclusion}
\label{section5}
In this paper, we present a detailed GC fluctuation theorem for the quantum systems described by the Floquet quantum master equations. Although this theorem is proved for systems that interact with one heat bath, its generalization to the case of multiple heat baths is straightforward. For instance, for the case of two heat baths, we have
\begin{eqnarray}
\label{extendedGCTtwoheatbath}
\frac{p(j_1^+,j_1^-,j_2^+,j_2^-)}{p(-j_1^-,-j_1^+,-j_2^-,-j_2^+) } \asymp e^{t\beta_1 ( j_1^++j_1^-)+t\beta_2(j_2^++j_2^-) },
\end{eqnarray}
where $j_k^+$ and $j_k^-$ are the positive and negative heat currents of the quantum system released to the heat bath with the inverse temperature $\beta_k$ ($k$$=$$1,2$). Finally, if a quantum system contacts two heat baths at two different temperatures and can be described by a Lindblad quantum master equation, the system is able to evolve into a nonequilibrium steady state without external driving fields.
% Editor: Please ensure that the intended meaning has been maintained in the edits of the previous sentence.
In such a situation, by carrying out the same argument presented here, we can prove that the fluctuation theorem~(\ref{extendedGCTtwoheatbath}) is still true.

\begin{acknowledgments}
This work was supported by the National Science Foundation of China under Grants No. 11174025 and No. 11575016. We also appreciate the support of the CAS Interdisciplinary Innovation Team, No. 2060299.
\end{acknowledgments}

\appendix

\end{document}